\documentclass[a4paper,twoside]{article}
\usepackage[dvips]{graphicx}
\usepackage{latexsym,epsf}
\usepackage{multicol}
\usepackage[centertags]{amsmath}
\usepackage [cp1251]{inputenc}
\usepackage [ukrainian, russian, english]{babel}
\usepackage {fancyhdr}

\newcommand{\bs}{\begin{sloppypar}} \newcommand{\es}{\end{sloppypar}}
\def\beq{\begin{eqnarray}} \def\eeq{\end{eqnarray}}
\def\beqstar{\begin{eqnarray*}} \def\eeqstar{\end{eqnarray*}}
\newcommand{\bal}{\begin{align}}
\newcommand{\eal}{\end{align}}
\newcommand{\beqe}{\begin{equation}} \newcommand{\eeqe}{\end{equation}}
\newcommand{\p}[1]{(\ref{#1})}
\newcommand{\tbf}{\textbf}

\fancyfootoffset{1pt}%
\makeatletter
\renewcommand{\ps@plain}{
\renewcommand{\@oddhead}{}
\renewcommand{\@evenhead}{}
\renewcommand{\@oddfoot}{\hfil \thepage}
\renewcommand{\@evenfoot}{\thepage \hfil \hfil}}
\makeatother 

\makeatletter
\renewcommand{\@biblabel}[1]{#1.\hfill}
\makeatother
\title{\textbf{\Large IS A FIELD-INDUCED FERROMAGNETIC PHASE TRANSITION IN THE MAGNETAR
CORE ACTUALLY POSSIBLE?}}
\author{\textbf{\textit{A.A. Isayev$\,^{1,\,2}$\footnote{\normalfont
 E-mail address: isayev@kipt.kharkov.ua
}, J. Yang$\,^3$\footnote{\normalfont E-mail address:
jyang@ewha.ac.kr}  }}
\\
\emph{\small $^1$National Science Center "Kharkov Institute of
Physics and Technology",  61108,  Kharkov, Ukraine}\\
\emph{\small $^2$Kharkov National University, Svobody Sq., 4,
Kharkov, 61077, Ukraine}\\
\emph{\small $^3$Department  of Physics and the Institute for the
Early Universe,}\\ \emph{\small
 Ewha Womans University, Seoul 120-750, Korea}\\
}
\begin{document}
\selectlanguage{english}
\date{}
\maketitle

\thispagestyle{fancy}
\begin{center}
\begin{minipage}{165mm}
{\small
%----------------------   Abstract  --------------------------------%
Spin polarized states in dense neutron matter with  BSk20 Skyrme
force are considered in magnetic fields up to $10^{20}$~G. It is
shown that the appearance of the longitudinal instability in a
strong magnetic field prevents the formation of a fully spin
polarized state in neutron matter, and only the states with moderate
spin polarization can be developed. }
\par \vspace{1ex}
%-------------------------------------------------------------------%
\end{minipage}
\end{center}
%--------------------  Text of the article  ------------------------%
\begin{multicols}{2}
\begin{center}
\textbf{\textsc{1. INTRODUCTION. BASIC EQUATIONS }}
\end{center}

\label{I}Magnetars are strongly magnetized neutron stars with
emissions powered by the dissipation of magnetic energy.
 The magnetic field strength at the surface
of a magnetar is  of about $10^{14}$-$10^{15}$~G. Such huge magnetic
fields can be inferred from observations of magnetar periods and
spin-down rates, or from hydrogen spectral lines. In the interior of
a magnetar the magnetic field strength could reach values up to
$10^{20}$~G~\cite{FIKPS}. Then the issue of interest is the behavior
of
 neutron star matter, which further will be approximated by pure neutron matter,
  in a strong magnetic
field~\cite{IY09}. In particular,  a scenario is possible in which a
field-induced ferromagnetic phase transition occurs in the magnetar
core. This idea was explored in the recent research~\cite{BRM},
where it was shown  that a fully spin polarized state in neutron
matter  could be formed in the magnetic field larger than
$10^{19}$~G. Note, however, that   the breaking of the ${\cal O}(3)$
rotational symmetry in such ultrastrong magnetic fields results in
the
 anisotropy of the total pressure, having a smaller value along than
perpendicular to the field direction~\cite{FIKPS,Kh}. The possible
outcome could be the gravitational collapse of a magnetar along the
magnetic field, if the magnetic field strength is large enough.
  Thus, exploring the possibility of a
field-induced ferromagnetic phase transition in neutron matter in a
strong magnetic field, the effect of the pressure anisotropy has to
be taken into account because this kind of instability could prevent
the formation of a fully polarized state in neutron matter. In the
 present
 study, we determine   thermodynamic quantities
 of strongly magnetized
   neutron matter
 taking into account this effect.

% \eject
%\newpage
%\vspace{3mm}

Let us stop on the basic equations of the theory. The normal
(nonsuperfluid) states of neutron matter are described
  by the normal distribution function of neutrons $f_{\kappa_1\kappa_2}=\mbox{Tr}\,\varrho
  a^+_{\kappa_2}a_{\kappa_1}$, where
$\kappa\equiv({\bf{p}},\sigma)$, ${\bf p}$ is momentum, $\sigma$ is
the projection of spin on the third axis, and $\varrho$ is the
density matrix of the system~\cite{IY,I06}.  The energy of the
system is specified as a functional of the distribution function
$f$, $E=E(f)$, and determines the single particle
energy~\cite{AIP,AIPY}
 \begin{eqnarray}
%$\delta E={\rm tr}\, \varepsilon(f)\delta f,$%
\varepsilon_{\kappa_1\kappa_2}(f)=\frac{\partial E(f)}{\partial
f_{\kappa_2\kappa_1}}.\label{1} \end{eqnarray}
%$\mathrm{tr}...$
%being the trace in the space of $\kappa$ variables.
The
self-consistent matrix equation for determining the distribution
function $f$ follows from the minimum condition of the thermodynamic
potential~\cite{AIP} and is
  \begin{align}\label{2}
 f&=\left\{\mbox{exp}(Y_0\varepsilon+Y_i\cdot \mu_n\sigma_i+
Y_4)+1\right\}^{-1}\\ &\equiv
\left\{\mbox{exp}(Y_0\xi)+1\right\}^{-1}.\nonumber \end{align} Here
the quantities $\varepsilon, Y_i$ and $Y_4$ are matrices in the
space of $\kappa$ variables, with
$\bigl(Y_{i,4}\bigr)_{\kappa_1\kappa_2}=Y_{i,4}\delta_{\kappa_1\kappa_2}$,
$Y_0=1/T$, $Y_i=-H_i/T$ and $ Y_{4}=-\mu_0/T$  being
 the Lagrange multipliers, $\mu_0$ being the chemical
potential of  neutrons, and $T$  the temperature. In Eq.~\p{2},
$\mu_n=-1.9130427(5)\mu_N$ is the neutron magnetic moment ($\mu_N$
being the nuclear  magneton), $\sigma_i$ are the Pauli matrices.

Further it will be assumed that the third axis is directed along the
external magnetic field $\bf{H}$. Given the possibility for
alignment of neutron spins along or opposite to the magnetic field
$\bf H$, the normal distribution function of neutrons and the matrix
quantity  $\xi$ (which we will also call a single particle energy)
can be expanded in the Pauli matrices $\sigma_i$ in spin
space%~\cite{AIP}
\begin{align} f({\bf p})&= f_{0}({\bf
p})\sigma_0+f_{3}({\bf p})\sigma_3,\label{7.2}\\
\xi({\bf p})&= \xi_{0}({\bf p})\sigma_0+\xi_{3}({\bf p})\sigma_3.
 %\nonumber
\end{align}

% \bal
%\omega_{\pm}&=\xi_{0}\pm\xi_{3},\label{omega}\\
%\xi_{0}&=\varepsilon_{0}-\mu_{0},\;
%\xi_{3}=-\mu_nH+\varepsilon_{3}.\nonumber\end{align}

The distribution functions $f_0,f_3$ satisfy the normalization
conditions
\begin{align} \frac{2}{\cal
V}\sum_{\bf p}f_{0}({\bf p})&=\varrho,\label{3.1}\\
\frac{2}{\cal V}\sum_{\bf p}f_{3}({\bf
p})&=\varrho_\uparrow-\varrho_\downarrow\equiv\Delta\varrho.\label{3.2}
 \end{align}
 Here $\varrho=\varrho_{\uparrow}+\varrho_{\downarrow}$ is the total density of
 neutron matter, $\varrho_{\uparrow}$ and $\varrho_{\downarrow}$  are the neutron number densities
 with spin up and spin down,
 respectively. The
quantity $\Delta\varrho$  may be regarded as the neutron spin order
parameter which  determines the magnetization of the system $M=\mu_n
\Delta\varrho$.
 The magnetization may
contribute to the internal magnetic field $\tbf{B}=\tbf{H}+4\pi
\tbf{M}$. However, we will assume, analogously to the previous
studies~\cite{IY09}, that, because of the tiny value of the neutron
magnetic moment, the contribution of the magnetization to the inner
magnetic field $\bf{B}$ remains small for all relevant densities and
magnetic field strengths, and, hence, $ \bf{B}\approx \bf{H}.$ In
order to get the self--consistent equations for the components of
the single particle energy, one has to set the energy functional of
the system. It represents the sum of the matter and field energy
contributions
\begin{equation}\label{en}
E(f,H)=E_m(f)+\frac{H^2}{8\pi}{\cal V}.
\end{equation}
The matter energy is the sum of the kinetic and Fermi-liquid
interaction energy terms~\cite{IY,I06}
\begin{align} E_m(f)&=E_0(f)+E_{int}(f),\label{enfunc} \\
{E}_0(f)&=2\sum\limits_{ \bf p}^{}
\underline{\varepsilon}_{\,0}({\bf p})f_{0}({\bf p}),\nonumber
\\ {E}_{int}(f)&=\sum\limits_{ \bf p}^{}\{
\tilde\varepsilon_{0}({\bf p})f_{0}({\bf p})+
\tilde\varepsilon_{3}({\bf p})f_{3}({\bf p})\},\nonumber \end{align}
where
\begin{align}\tilde\varepsilon_{0}({\bf p})&=\frac{1}{2\cal
V}\sum_{\bf q}U_0^n({\bf k})f_{0}({\bf
q}),\;{\bf k}=\frac{{\bf p}-{\bf q}}{2}, \label{ve0}\\
\tilde\varepsilon_{3}({\bf p})&=\frac{1}{2\cal V}\sum_{\bf
q}U_1^n({\bf k})f_{3}({\bf q}). \label{ve3}%\nonumber
\end{align}
Here  $\underline\varepsilon_{\,0}({\bf p})=\frac{{\bf
p}^{\,2}}{2m_{0}}$ is the free single particle spectrum, $m_0$ is
the bare mass of a neutron, $U_0^n({\bf k}), U_1^n({\bf k})$ are the
normal Fermi liquid (FL) amplitudes, and
$\tilde\varepsilon_{0},\tilde\varepsilon_{3}$ are the FL corrections
to the free single particle spectrum. Taking into account
Eqs.~\p{1},\p{2} and \p{enfunc},  expressions for the components of
the single particle energy read \bal\xi_{0}({\bf
p})&=\underline{\varepsilon}_{\,0}({\bf
p})+\tilde\varepsilon_{0}({\bf p})-\mu_0,\; \xi_{3}({\bf
p})=-\mu_nH+\tilde\varepsilon_{3}({\bf p}).\label{14.2}
\end{align}

In Eqs.~\p{14.2}, the quantities
$\tilde\varepsilon_{0},\tilde\varepsilon_{3}$ are the functionals of
the distribution functions $f_0,f_3$ which, using Eqs.~\p{2} and
\p{7.2}, can be expressed, in turn, through the quantities $\xi$:
\begin{align}
f_{0}&=\frac{1}{2}\{n(\omega_{+})+n(\omega_{-}) \},\label{2.4}
 \\
f_{3}&=\frac{1}{2}\{n(\omega_{+})-n(\omega_{-})\},\label{2.5}
 \end{align} where
 %\begin{align*}
 $$
    n(\omega_\pm)=\{\exp(Y_0\omega_\pm)+1\}^{-1},\quad
\omega_{\pm}=\xi_{0}\pm\xi_{3}.$$
%\end{align*}

Thus, Eqs.~\p{14.2}--\p{2.5} form the self-consistency equations for
the components of the single particle energy, which should be solved
jointly with the normalization conditions~\p{3.1}, \p{3.2}.

The pressures (longitudinal and transverse with respect to the
direction of the magnetic field) in the system are related to the
diagonal elements of the stress tensor whose explicit expression
reads~\cite{LLP}

\begin{equation}\label{sigma}
    \sigma_{ik}=\biggl[\tilde{ \textsl{f}}-\varrho\biggl(\frac{\partial
    \tilde{ \textsl{f}}}{\partial \varrho}\biggr)_{{\bf
    H},T}\biggr]\delta_{ik}+\frac{H_iB_k}{4\pi}.
\end{equation}
Here
\begin{equation}\label{Ft}
\tilde{ \textsl{f}}=\textsl{f}_H-\frac{H^2}{4\pi},%\equiv
%\mathfrak{f}-{\bf HM}-\frac{{\bf H}^2}{4\pi},
\end{equation}
$\textsl{f}_H=\frac{1}{\cal V}(E-TS)-\tbf{HM}$ is the Helmholtz free
energy density. For the isotropic medium, the stress
tensor~\p{sigma} is symmetric. The transverse  $p_{t}$  and
longitudinal $p_{l}$  pressures are determined from the formulas
\begin{equation*}
p_{t}=-\sigma_{11}=-\sigma_{22},\; p_{l}=-\sigma_{33}.
\end{equation*}
At zero temperature, using Eqs.~\p{en}, \p{sigma}, one can get the
approximate expressions
\begin{align}\label{press}
    %p_t&=\varrho^2\frac{\partial(
    %e_m/\varrho)}{\partial \varrho}+\frac{H^2}{8\pi},\\
    %p_l&=\varrho^2\frac{\partial(
    %e_m/\varrho)}{\partial \varrho}-\frac{H^2}{8\pi},
p_t&=\varrho\Bigl(\frac{\partial
    e_m}{\partial \varrho}\Bigr)_{H}-e_m+\frac{H^2}{8\pi},\\
    p_l&=\varrho\Bigl(\frac{\partial
    e_m}{\partial \varrho}\Bigr)_{H}-e_m-\frac{H^2}{8\pi},
    \end{align}
where $e_m$ is the  matter energy density, and we disregarded the
terms proportional to $M$.  In ultrastrong magnetic fields, the
quadratic on the magnetic field term (the Maxwell term) will be
dominating, leading to increasing the transverse pressure and to
decreasing the longitudinal pressure. Hence, at some critical
magnetic field, the longitudinal pressure vanishes, resulting in the
longitudinal instability  of neutron matter. The question then is:
What is the magnitude of the critical field and
 the corresponding maximum degree of spin polarization in neutron
matter? %\eject
\vspace{6mm}
\begin{center}
\textbf{\textsc{2. EOS  OF DENSE NEUTRON MATTER IN A STRONG MAGNETIC
FIELD}}
\end{center}

In numerical calculations, we utilize the BSk20 Skyrme
force~\cite{GCP} constrained such as to avoid the spontaneous spin
instability of neutron matter at densities beyond the nuclear
saturation density and to reproduce a microscopic EoS of
nonpolarized neutron matter. Expressions for the normal FL
amplitudes in Eqs.~\p{ve0},\p{ve3} in terms of the parameters of the
Skyrme interaction are given in Ref.~\cite{IY10a}. Now we present
the results of the numerical solution  of the self-consistency
equations. Fig.~1 shows the spin polarization parameter $
    \Pi=\frac{\Delta\varrho}{\varrho}
$ of neutron matter as a function of the magnetic field $H$ at two
different values of the neutron matter density, $\varrho=3\varrho_0$
and $\varrho=4\varrho_0$, which can be relevant for the magnetar
core. It is seen that the impact of the magnetic field remains small
up to the field strength $10^{17}$~G.  The larger the density is,
the smaller the effect produced by the magnetic field on spin
polarization of neutron matter.

\begin{center}
\begin{minipage}{80 mm}
\includegraphics[width=\textwidth]{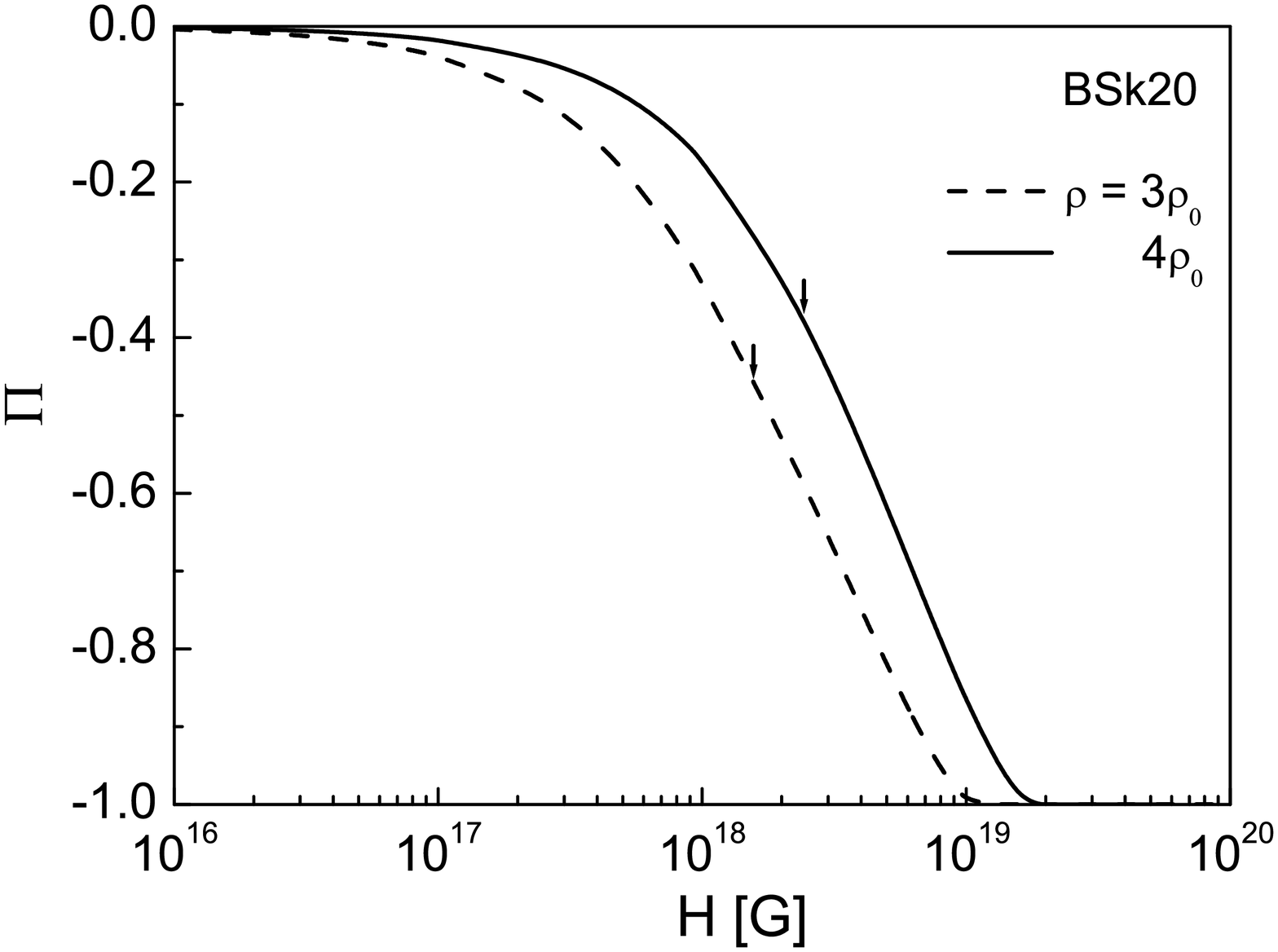}
\emph{\textbf{Fig.1.}} {\emph{Neutron spin polarization parameter as
a function of the magnetic field $H$ for the Skyrme force BSk20 at
zero temperature and fixed values of the density,
$\varrho=3\varrho_0$ and $\varrho=4\varrho_0$. The vertical arrows
indicate the maximum magnitude of spin polarization attainable  at
the given density, see further details in the text.  }}\\
\end{minipage}
\end{center}

 At the magnetic
field $H=10^{18}$~G, usually considered as the maximum magnetic
field strength in the core of a magnetar (according to a scalar
virial theorem, see Ref.~\cite{FIKPS} and references therein), the
magnitude of the spin polarization parameter doesn't exceed $33\%$
at $\varrho=3\varrho_0$ and $18\%$ at $\varrho=4\varrho_0$. However,
the situation changes if the larger magnetic fields are allowable:
With further increasing the magnetic field strength, the magnitude
of the spin polarization parameter increases till it reaches the
limiting value $\Pi=-1$, corresponding to a fully spin polarized
state. For example, this happens at $H\approx 1.25\cdot 10^{19}$~G
for $\varrho=3\varrho_0$ and at $H\approx 1.98\cdot 10^{19}$~G for
$\varrho=4\varrho_0$, i.e., certainly, for magnetic fields larger
than $10^{19}$~G. Nevertheless, we should check whether the
formation of a fully spin polarized state in a strong magnetic field
is actually possible by calculating the anisotropic pressure in
dense neutron matter. The meaning of the vertical arrows in Fig.~1
is explained later in the text.

\begin{center}
\begin{minipage}{80 mm}
%\begin{figure}[tb]
%\begin{center}
\includegraphics[width=\textwidth]{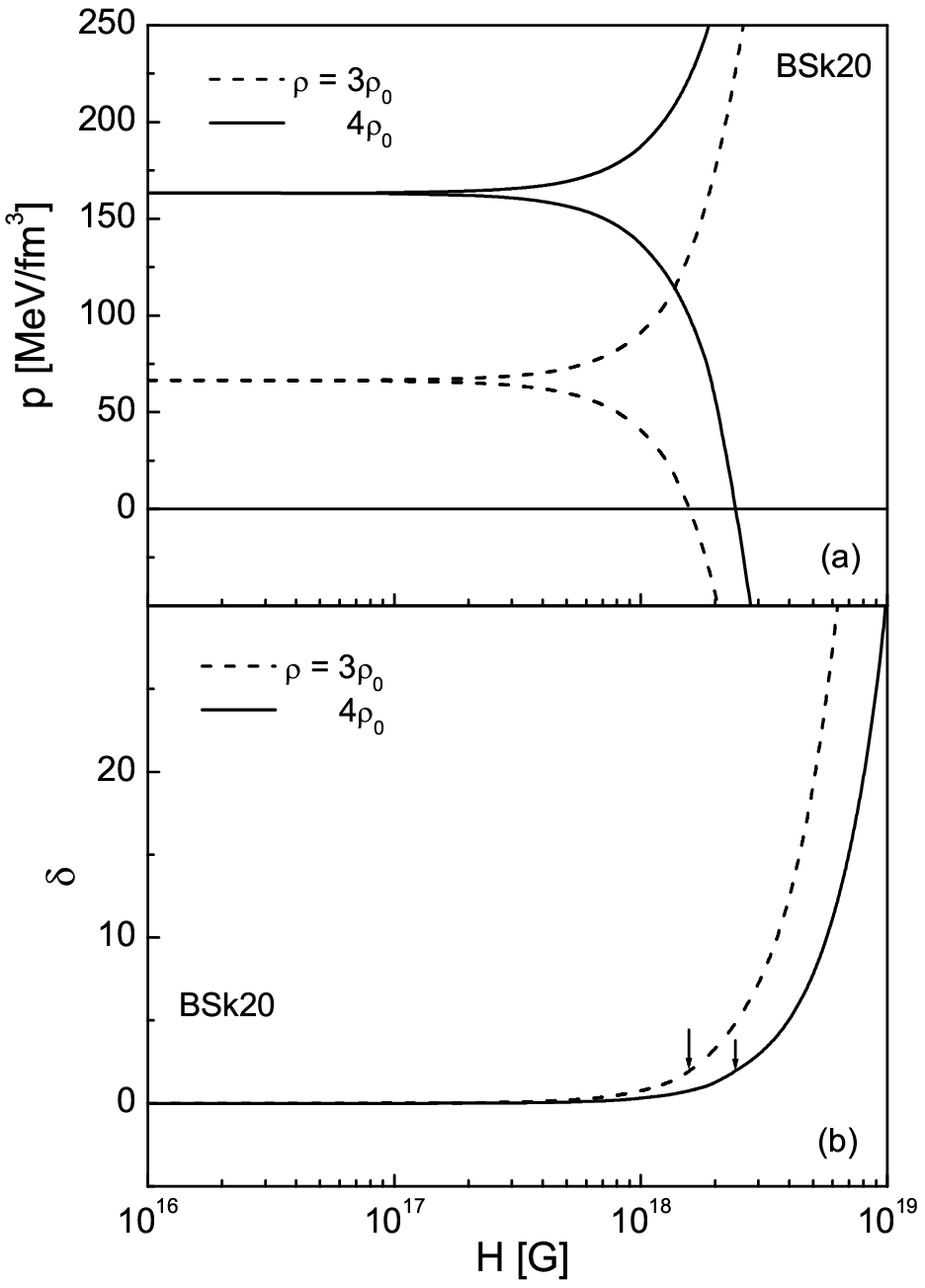}
%\end{center}
\emph{\textbf{Fig.2.}} {\emph{(a) Pressures, longitudinal
(descending branches) and transverse (ascending branches),
 as  functions of the magnetic field $H$ for the
Skyrme force BSk20 at zero temperature and fixed values of the
density, $\varrho=3\varrho_0$  and $\varrho=4\varrho_0$. (b) Same as
in the top panel but for the normalized difference between the
transverse and longitudinal pressures. }}\\
%\end{figure}
\end{minipage}
\end{center}

Fig.~2a shows the pressures (longitudinal and transverse) in neutron
matter as functions of the magnetic field $H$ at the same densities,
$\varrho=3\varrho_0$ and $\varrho=4\varrho_0$. First, it is clearly
seen that up to some threshold  magnetic field the difference
between transverse and longitudinal pressures is unessential that
corresponds to the isotropic regime. Beyond this threshold magnetic
field strength, the anisotropic regime holds for which the
transverse pressure increases with $H$ while the longitudinal
pressure decreases.  The longitudinal pressure vanishes at some
critical magnetic field $H_c$ marking the onset of the longitudinal
collapse of a neutron star. For example, $H_c\approx1.56\cdot
10^{18}$~G at $\varrho=3\varrho_0$ and $H_c\approx2.42\cdot
10^{18}$~G at $\varrho=4\varrho_0$. In all cases under
consideration, this critical value doesn't exceed $10^{19}$~G.

 The magnitude of the spin
polarization parameter $\Pi$ cannot also exceed some limiting value
corresponding to the critical field $H_c$. These maximum values of
the $\Pi$'s magnitude are shown in Fig.~1 by the vertical arrows. In
particular, $\Pi_c\approx-0.46$ at $\varrho=3\varrho_0$ and
$\Pi_c\approx-0.38$ at $\varrho=4\varrho_0$. As can be inferred from
these values, the appearance of the negative longitudinal pressure
in an ultrastrong magnetic field prevents the formation of a fully
spin polarized state in the core of a magnetar. Therefore, only the
onset of a field-induced ferromagnetic phase transition, or its near
vicinity, can be catched under increasing the magnetic field
strength in dense neutron matter. A complete spin polarization
 in the magnetar
core is not allowed by the appearance of the negative pressure along
the direction of the magnetic field, contrary to the conclusion of
Ref.~\cite{BRM} where the pressure anisotropy in a strong magnetic
field was disregarded.

Fig.~2b shows the difference between the transverse and longitudinal
pressures normalized to the value of the pressure $p_0$ in the
isotropic regime (which corresponds to the weak field limit with
$p_l=p_t=p_0$) being $\delta=\frac{p_{t}-p_{l}}{p_0}. $ Applying for
the transition from the isotropic regime to the anisotropic one the
criterion $\delta\simeq 1$, the transition occurs at the threshold
field $H_{th}\approx 1.15\cdot 10^{18}$~G for  $\varrho=3\varrho_0$
and $H_{th}\approx1.83\cdot 10^{18}$~G for $\varrho=4\varrho_0$. In
all cases under consideration, the threshold field $H_{th}$ is
larger than $10^{18}$~G, and, hence, the isotropic regime holds for
the fields up to $10^{18}$~G.   The vertical arrows in Fig.~2b
indicate the points corresponding to the onset of the longitudinal
instability in neutron matter.  The maximum allowable normalized
splitting of the pressures corresponding to the critical field $H_c$
is $\delta\sim 2$.

\begin{center}
\begin{minipage}{80 mm}
\includegraphics[width=\textwidth]{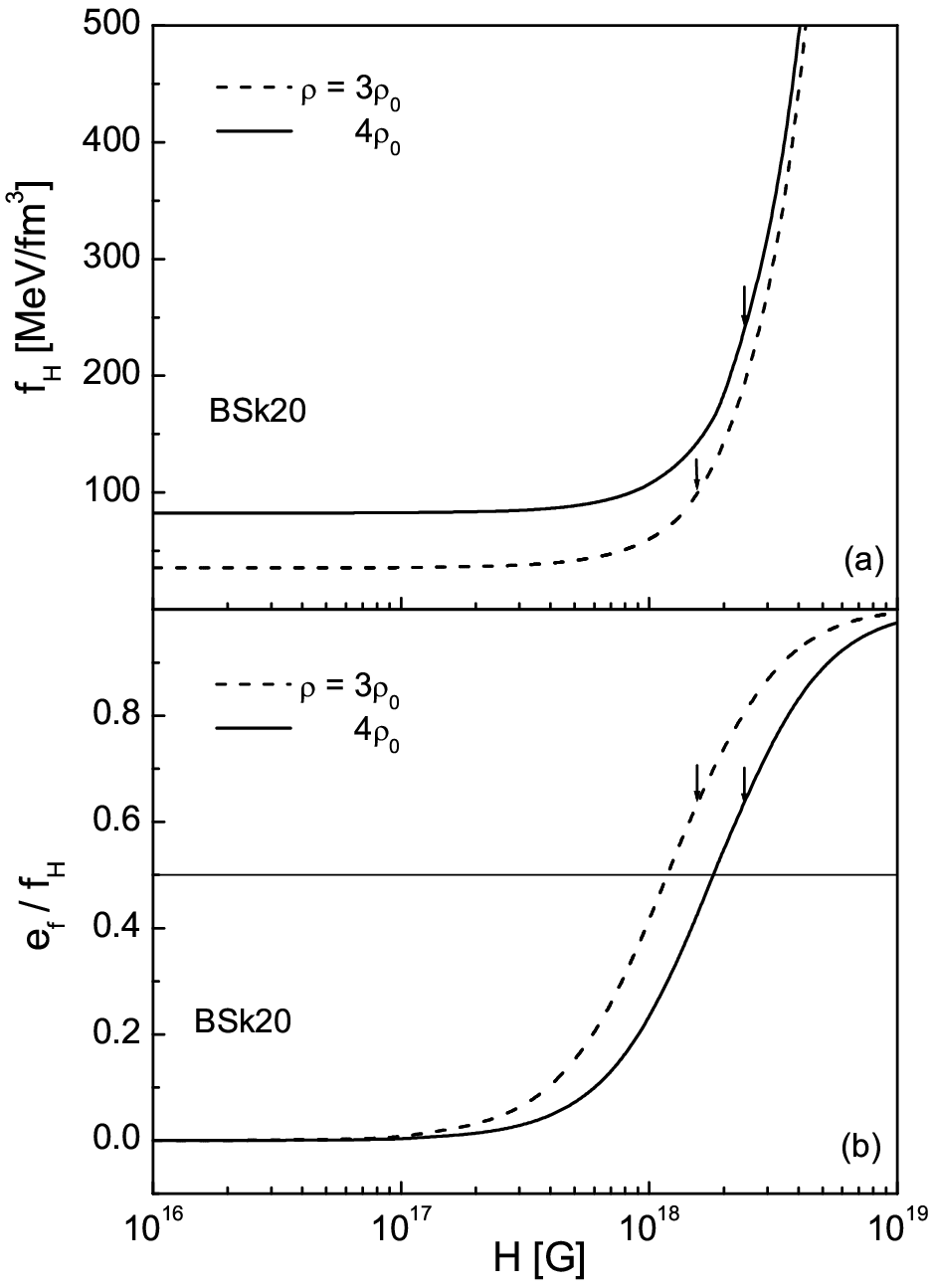}
\emph{\textbf{Fig.3.}} {\emph{Same as in Fig.~2 but for: (a)  the
Helmholtz free energy density of the system;   (b)  the ratio of the
magnetic field energy density to the Helmholtz free energy density
of the system.  }}\\
\end{minipage}
\end{center}

Fig.~3a shows the Helmholtz free energy density of the system as a
function of the magnetic field $H$. It is seen that the magnetic
fields up to $H \sim10^{18}$~G have practically small effect on the
Helmholtz free energy density $\textsl{f}_H$, but beyond this field
strength the contribution of the magnetic field energy to the free
energy $\textsl{f}_H$ rapidly increases with $H$. However, this
increase is limited by the values of the critical magnetic field
corresponding to the onset of the longitudinal instability in
neutron matter. The respective points on the curves are indicated by
the vertical arrows.

\begin{center}
\begin{minipage}{80 mm}
\includegraphics[width=\textwidth]{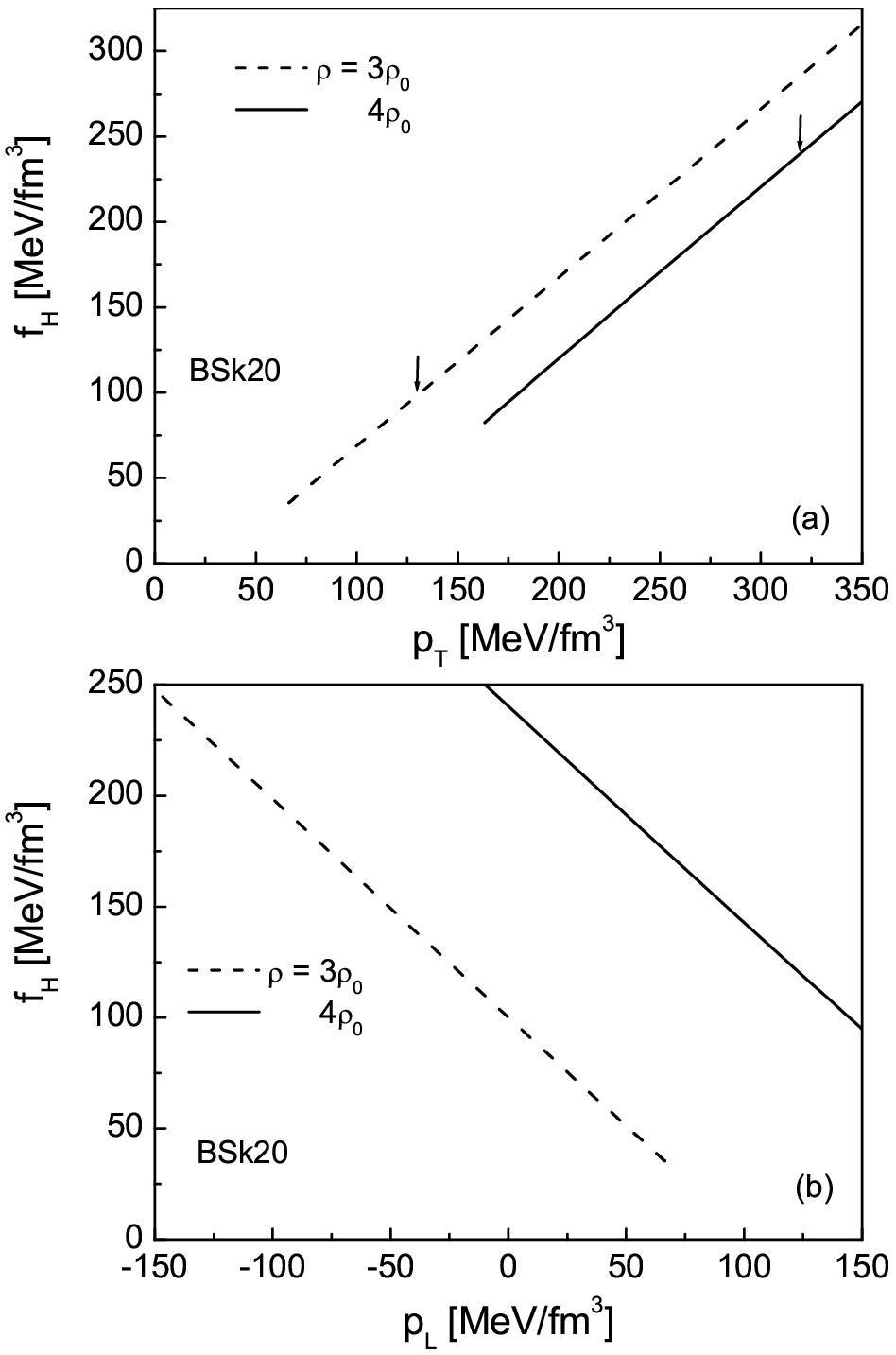}
\emph{\textbf{Fig.4.}} {\emph{The Helmholtz free energy density of
the system  as  a function of: (a) the transverse pressure $p_t$,
(b) the longitudinal  pressure $p_l$  for the Skyrme force BSk20 at
zero temperature and fixed values of the density,
$\varrho=3\varrho_0$ and $\varrho=4\varrho_0$.  }}\\
\end{minipage}
\end{center}

Fig.~3b shows the ratio of the magnetic field energy density
$e_f=\frac{H^2}{8\pi}$ to the Helmholtz free energy density at the
same assumptions as in Fig.~2. The intersection points of the
respective curves in this panel  with the line
$e_f/\textsl{f}_H=0.5$ correspond to the magnetic fields at which
the matter and field contributions to the Helmholtz free energy
density are equal. This happens at $H\approx1.18\cdot10^{18}$~G  for
$\varrho=3\varrho_0$, and at $H\approx1.81\cdot10^{18}$~G  for
$\varrho=4\varrho_0$. These values are quite close to the respective
values of the threshold field $H_{th}$, and, hence, the transition
to the anisotropic regime occurs at the magnetic field strength at
which the field and matter contributions to the Helmholtz free
energy density become equally important. It is also seen from
Fig.~3b that in all cases when the longitudinal instability occurs
in the magnetic field $H_c$ the contribution of the magnetic field
energy density to the Helmholtz free energy density of the system
dominates over the matter contribution.

Because of the pressure anisotropy, the EoS of neutron matter in a
strong magnetic field is also anisotropic. Fig.~4 shows the
dependence of the Helmholtz free energy density $\textsl{f}_H$ of
the system on the transverse pressure (top panel) and on the
longitudinal pressure (bottom panel) at the same densities
considered above.
 Since in an ultrastrong magnetic field
 the dominant Maxwell  term enters the pressure $p_t$ and free energy
density $\textsl{f}_H$ with  positive sign and the pressure $p_l$
with negative sign, the free energy density $\textsl{f}_H$ is the
increasing function of $p_t$ and decreasing function of $p_l$.  In
the bottom panel, the physical region corresponds to the positive
values of the longitudinal pressure.

The obtained results can be of importance in the structure studies
of magnetars. It would be also of interest to extend this research
to finite temperatures relevant for proto-neutron stars which can
lead to a number of interesting effects, such as, e.g., an unusual
behavior of the entropy of a spin polarized state~\cite{I3,I4}.

J.Y. was supported by grant 2010-0011378 from Basic Science Research
Program through NRF of Korea funded by MEST and by  grant R32-10130
from WCU project of MEST and NRF.

 \vspace{2mm}
%--------------------   Bibliography  ------------------------------------%
\begin{center}

\end{center}
\end{multicols}

\begin{thebibliography}{99}



\bibitem{FIKPS} E. J. Ferrer, V. de la Incera, J. P. Keith, I. Portillo, and P. L.
Springsteen. Equation of state of a dense and magnetized fermion
system //  \textit{Phys. Rev. C} 2010, 82, 065802, p. 15.



  \bibitem{IY09} A.A.  Isayev, and  J. Yang.   Spin-polarized states in neutron
matter in a strong magnetic field // \textit{Phys.  Rev.  C} 2009,
80, 065801, p. 7.



\bibitem{BRM} G. H. Bordbar, Z. Rezaei, and A. Montakhab.  Investigation of the field-induced
ferromagnetic phase transition in spin-polarized neutron matter: A
lowest order constrained variational approach //   \textit{Phys.
Rev. C} 2011, 83, 044310, p. 7.

\bibitem{Kh} V. R. Khalilov. Macroscopic effects in cold magnetized nucleons and
electrons with anomalous magnetic moments // \textit{Phys. Rev. D}
2002, 65, 056001, p. 6.





\bibitem{IY}  A.A.  Isayev, and  J. Yang. Spin polarized states in strongly asymmetric nuclear matter
//   \textit{Phys.  Rev.  C} 2004,  69, 025801, p. 8.


\bibitem{I06}
   A.A. Isayev.  Spin ordered phase transitions in isospin asymmetric
                  nuclear matter // \textit{Phys.  Rev.  C} 2006,  74,
                  057301, p. 4.

\bibitem{AIP} A. I. Akhiezer, A. A. Isayev, S. V. Peletminsky, A. P. Rekalo, and
A. A. Yatsenko. On a theory of superfluidity of nuclear matter based
on the Fermi-liquid approach // \textit{JETP} 1997,  85, 1-12.

\bibitem{AIPY} A. I. Akhiezer, A. A. Isayev, S. V. Peletminsky, and
A. A. Yatsenko. Multi-gap superfluidity in nuclear matter //
\textit{Phys. Lett. B} 1999,  451, 430–436.





\bibitem{LLP}L. D. Landau, E. M. Lifshitz, and L. P. Pitaevskii.
\emph{Electrodynamics of Continuous Media}, New York, Pergamon,
1984, 2nd ed.

\bibitem{GCP} S. Goriely, N. Chamel,  and J. M. Pearson.
Further explorations of Skyrme-Hartree-Fock-Bogoliubov mass
formulas. XII. Stiffness and stability of neutron-star matter //
\textit{Phys.  Rev.  C} 2010, 82, 035804, p. 18.



\bibitem{IY10a} A.A.  Isayev, and  J. Yang.
Phase transition to the state with nonzero average helicity  in
dense neutron matter // \textit{ JETP Lett.} 2010,  92, p. 783-787.

\bibitem{I3}
A.A. Isayev. Finite temperature effects in antiferromagnetism of
nuclear matter // \textit{Phys.  Rev.  C} 2005, 72, 014313, p. 4.

\bibitem{I4} A.A. Isayev. Unusual temperature behavior of the entropy of the antiferromagnetic spin state
in nuclear matter with an effective finite range interaction //
\textit{Phys.  Rev.  C} 2007, 76, 047305, p.~4.


\end{thebibliography}
\end{document}